\documentclass[12pt]{article}

\usepackage[titletoc]{appendix}

\usepackage[margin={1.0in}]{geometry}

\usepackage{booktabs}
\usepackage{threeparttablex}

\usepackage{subfig}

\usepackage{setspace}

\usepackage{natbib}

\usepackage{amsmath}
\usepackage{amssymb}
\usepackage{mathtools}

\usepackage{authblk}
\usepackage{dsfont}

\usepackage{anyfontsize}
\usepackage{lmodern}\usepackage{anyfontsize}\usepackage[sc]{mathpazo}
\usepackage{courier}
\usepackage[utf8]{inputenc}
\usepackage[T1]{fontenc}
\usepackage{mathrsfs}
\usepackage{amsthm}
\usepackage{bbm}

\usepackage{float}\usepackage{multirow}
\usepackage{enumerate}
\usepackage{hyperref}
\usepackage[nameinlink]{cleveref}

\hypersetup{
 colorlinks=true, linktoc=all, linkcolor=black, urlcolor  = blue,
 citecolor = black,
    pdftitle={Bank Failures: The Roles of Solvency and Liquidity},
    pdfauthor={Sergio Correia, Stephan Luck, Emil Verner}
}

\usepackage[nottoc,notlot,notlof]{tocbibind}

\hypersetup{pdfstartview={Fit}}\RequirePackage[font=small,format=plain,labelfont=bf,textfont=it]{caption}

\newtheorem{hypothesis}{Hypothesis}

\makeatletter
\@ifundefined{tabnote}{
  \newenvironment{tabnote}[1][\linewidth]{    \par\vspace{.5\baselineskip}    \noindent\begin{minipage}{#1}\footnotesize
  }{    \end{minipage}\par\vspace{.25\baselineskip}  }}{}
\makeatother

\let\maybescriptsize\scriptsize

\begin{document}

\title{\vspace{0cm} \LARGE \bf Bank Failures: \\The Roles of Solvency and Liquidity}

\author{\vspace{.5cm} Sergio Correia, Stephan Luck, and Emil Verner\textsuperscript{*} }

\date{ \today }

\pagenumbering{gobble}
\maketitle

\begin{abstract}
Bank failures can stem from runs on otherwise solvent banks or from losses that render banks insolvent, regardless of withdrawals. Disentangling the relative importance of liquidity and solvency in explaining bank failures is central to understanding financial crises and designing effective financial stability policies. This paper reviews evidence on the causes of bank failures. Bank failures---both with and without runs---are almost always related to poor fundamentals. Low recovery rates in failure suggest that most failed banks that experienced runs were likely fundamentally insolvent.  Examiners' postmortem assessments also emphasize the primacy of poor asset quality and solvency problems. Before deposit insurance, runs commonly triggered the failure of insolvent banks. However, runs rarely caused the failure of strong banks, as such runs were typically resolved through other mechanisms, including interbank cooperation, equity injections, public signals of strength, or suspension of convertibility. We discuss the policy implications of these findings and outline directions for future research.

\end{abstract}

\let\oldthefootnote\thefootnote
\renewcommand{\thefootnote}{\fnsymbol{footnote}}
\footnotetext[1]{Correia: Federal Reserve Bank of Richmond, \href{mailto:sergio.correia@rich.frb.org}{sergio.correia@rich.frb.org}; Luck: Federal Reserve Bank of New York, \href{mailto:stephan.luck@ny.frb.org}{stephan.luck@ny.frb.org};
Verner: MIT Sloan School of Management and NBER \href{mailto:everner@mit.edu}{everner@mit.edu}.

We thank an anonymous referee and Thomas Eisenbach, Huberto Ennis, Sam Hanson, and Todd Keister for helpful comments. We also thank Trang Do for valuable research assistance. The opinions expressed in this article do not necessarily reflect those of the Federal Reserve Bank of New York or the Federal Reserve Bank of Richmond.

}

\let\thefootnote\oldthefootnote

\doublespacing
\pagenumbering{arabic}
\clearpage

\section{Introduction}

\begin{figure}
\includegraphics[width=1.0\linewidth]{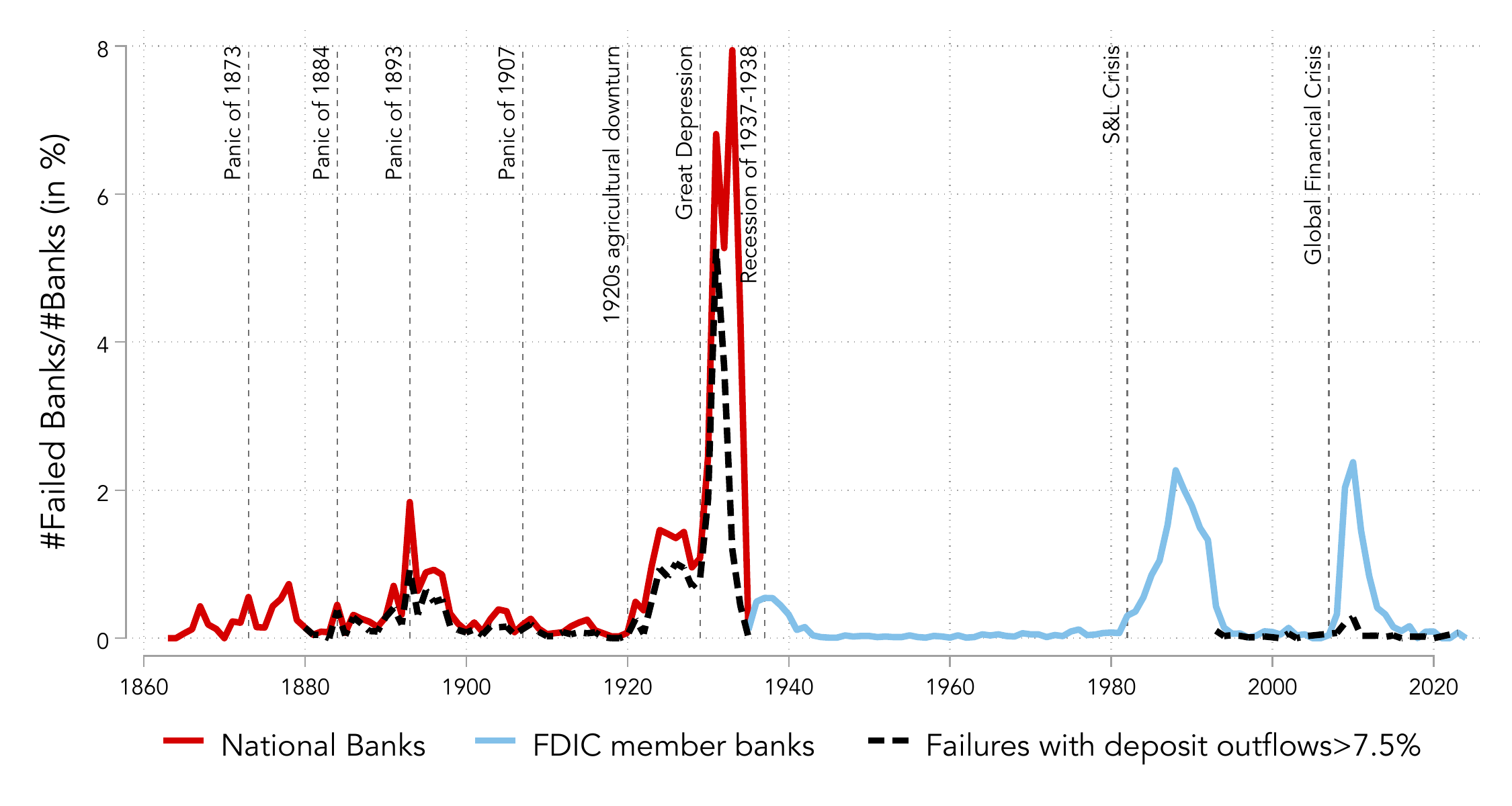}
\caption{\textbf{Bank Failures, With and Without Runs, 1863--2024 }\label{fig:failures}}
\begin{minipage}{\textwidth}
\footnotesize
Notes: This figure plots the rate of bank failures from 1863 to 2024, defined as 100 times the number of failed banks over the total number of banks. For 1863--1934, the failure rate is based on national banks, as reliable data on bank failures for state banks are not consistently available throughout the sample. For 1935--2024, the failure rate is based on Federal Deposit Insurance Corporation (FDIC) member banks. Failures with runs are defined failures in which deposit outflows exceed 7.5\% immediately before failure. Deposit outflows immediately before failure is based on the growth in deposits from the last Call Report before failure to the time of suspension. This information is available for 1880--1934 and 1992--2024.
\end{minipage}
\end{figure}

The U.S. banking system has repeatedly experienced major waves of bank failures.  \Cref{fig:failures} shows the rate of bank failures in the U.S. since 1863, highlighting spikes during the Panic of 1893, the 1920s agricultural downturn, the Great Depression, and the Global Financial Crisis. Each crisis reignites the age-old debate about whether bank failures are driven by illiquidity or insolvency. The debate reflects a theoretical ambiguity about why banks fail. In theory, bank failures can be the consequence of bank runs in which depositors collectively withdraw from otherwise solvent banks~\citep{Diamond1983} or from troubled but still solvent banks~\citep{Goldstein2005}. Alternatively, bank failures may primarily reflect insolvency due to poor fundamentals, regardless of deposit withdrawals.

Disentangling the roles of liquidity and solvency empirically is key to understanding bank failures and crises. It is also central to designing effective financial stability policies. Under the liquidity view, deposit insurance and public liquidity provision may suffice to avert costly bank failures. Under the solvency view, the emphasis shifts toward ensuring banks are well capitalized and restoring solvency after adverse shocks \citep{AdmatiHellwig2014}.

In this review, we synthesize the empirical evidence on illiquidity and insolvency as causes of bank failures and discuss what these findings imply for policies meant to address the incidence and consequences of financial crises. We narrow our focus along three dimensions. First, we focus on bank failures, defined as receiverships or other legal forms of bank resolution. The advantage of studying failures is that they are an informative proxy for bank distress that matters for understanding broader economic distress during crises \citep[see, e.g.,][]{Bernanke1983}. Failures are also objectively recorded in historical records. Nevertheless, we also touch on bank runs that did not result in failure. Second, we examine the long historical experience with bank failures. Historical settings are especially informative due to the absence of a public safety net to prevent bank runs. Third, our discussion primarily concerns the U.S. banking system, though we draw on international experience where relevant.\footnote{For aggregate international evidence on banking crises, see recent reviews by \citet{SufiTaylor2021FinancialCrisesSurvey} and  \citet{FrydmanXu2023}. For other perspectives on the roles of fundamentals and runs, see \citet{ennis2003economic}, \citet{GortonWinton2003}, and \citet{Goldstein2013}.}

We organize our discussion around a simple theory of bank runs and failures. The theory distinguishes between failures driven by insolvency or by runs on fundamentally solvent but potentially weak banks. The theory shows that these are related but distinct causes of bank failures. We use the theory to highlight several intuitive, testable hypotheses that we take to the data to distinguish between the two types of failures.

The theory predicts that both failures driven by insolvency and by runs are more likely in banks with weak fundamentals. The empirical evidence over the past 160 years of bank failures resoundingly supports this hypothesis. Bank failures are essentially always associated with weak fundamentals, proxied by bank capitalization, profitability, and reliance on expensive funding. Banks that fail with and without runs display similarly weak fundamentals. Runs do not trigger the failure of observably stronger banks. Moreover, weak fundamentals predict failures out-of-sample.

The strong relation between weak fundamentals and failures rejects the extreme view that runs frequently cause the failure of clearly healthy banks. However, it does not disentangle the roles of insolvency and runs in weak but solvent banks. While a bank's solvency status can be blurry, especially in the fog of a crisis, theory implies that recovery rates on bank assets provide additional insights into the relative importance of insolvency and illiquidity. In the historical U.S. banking system, recoveries were low, averaging 75\% of outstanding debt claims, including for failed banks that experienced a bank run before failure. Low recovery rates suggest that most failed banks, even those subject to runs, were likely insolvent, unless one assumes large value destruction caused by receivership. While bank runs were common in the pre-deposit insurance era, they were primarily triggers of failure for already insolvent banks.

Contemporary bank examiners' post-mortem assessments of the causes of bank failures provide complementary insights. These assessments tend to emphasize asset losses, economic conditions, and governance issues as explanations for why banks failed. In contrast, bank runs and heavy withdrawals were rarely mentioned as the primary causes of failure, with the notable exception of the Great Depression.

Why are bank runs on solvent banks not a common cause of bank failures? First, sleepy or inattentive depositors weaken strategic complementarities among depositors and reduce the scope for self-fulfilling runs to occur. The high empirical predictability of failures with runs, even in the absence of deposit insurance, suggests that depositors are often sleepy. This will tend to delay runs and increase the relative importance of solvency-driven failures. Second, mechanisms such as interbank cooperation, new equity injections, depositor relationships, examination, and suspension of convertibility allow illiquid but solvent banks to avoid costly fire sales. While these responses do not completely rule out the possibility that runs can cause the failure of solvent banks, we argue that they can substantially mitigate this risk, even absent government intervention.

Given that banking distress most commonly stems from fundamental solvency issues, the evidence suggests that policies focused on maintaining and restoring bank solvency, beyond merely providing liquidity support, are necessary for preventing the most adverse consequences of banking crises. However, more research is needed to reach definitive and general policy recommendations. Bank runs can have adverse consequences through channels other than bank failures. They can play a potentially outsized role during the most severe banking crises.

\section{Liquidity versus Solvency: Theory}
\label{sec:conceptual}

We begin with a simple illustrative framework that summarizes key themes from the theoretical literature on bank runs and failures~\citep[see, e.g.,][]{Diamond1983,Morris2003,Rochet2004,Goldstein2005}. We use this framework to derive empirically testable predictions. The framework distinguishes between fundamentally insolvent banks (that would have failed even absent a bank run) and banks whose failure was \textit{caused} by a run (that would have survived absent sudden withdrawals).

\paragraph*{Setup} There are two dates $t\in\{1,2\}$. At $t=1$, a bank holds a mix of liquid risk-free cash $C$ and illiquid risky loans $L$, financed by deposits $D$ and equity $E$. Total initial assets are thus $A=C+L=D+E$.
For simplicity, we assume that this initial capital structure is exogenous.\footnote{\citet{Diamond1983} show the conditions under which demand deposit contracts can insure depositors against idiosyncratic liquidity risk. A complementary rationale for the existence of short-term funding of banks and bank runs is provided by~\citet{Calomiris1991} and~\citet{Diamond2001}, who argue that demand deposit contracts are an instrument to discipline bank management.} Crucially, we assume that $D>C$, so the bank lacks the liquid funds to pay all depositors at once in the initial period. There is a continuum of risk-neutral depositors, each holding one unit of deposits. There is no discounting, deposits pay no interest, and the return on cash is zero. The gross return on loans is $\theta$, a deterministic parameter observable to all depositors at $t=1$.
Thus, at $t=2$ the bank's loan portfolio equals $\theta L$.

The only decision in the model is each depositor $i$'s choice at $t=1$ to maintain or withdraw their funds after observing $\theta$. This variable is $d_i\in\{0,1\}$, with $d_i=1$ indicating withdrawal. The bank can only serve depositors using its cash holdings. The loan portfolio is illiquid at $t=1$.  We assume sequential service of withdrawals. Depositors who withdraw while the bank is liquid receive full repayment; those who withdraw when it is illiquid receive only a pro rata share in receivership.

Denote the total amount of withdrawal requests in the initial period as $w=\int d_i \, di$. The key friction is that if the bank runs out of cash in this period, $w>C$, the bank is placed into receivership. In receivership, the loan portfolio's value is reduced to $(1-\rho)\theta L$. The reduction in value $\rho$ captures the notion that the value of loans may be higher when held within the bank than outside of the bank, e.g., due to the inalienability of a banker's human capital \citep[e.g.,][]{Diamond1984,HartMoore1994,Diamond2001}.

\paragraph*{Outcomes}

The bank is fundamentally insolvent when its assets cannot fully repay depositors, so it would default even without withdrawals. This occurs when $\theta<\theta^{Solvency}\equiv \frac{D-C}{L}$. In this case, it will be optimal for each depositor to withdraw even if others do not, as the sequential service constraint implies a chance of full repayment by arriving at the front of the queue. Thus, a bank run necessarily follows for $\theta<\theta^{Solvency}$.

When is a bank run an equilibrium even if the bank is not fundamentally insolvent and $\theta>\theta^{Solvency}$? Illiquidity occurs when withdrawal requests $w$ exceed available cash ($w>C$). In this case, a bank run can be an equilibrium if at $t=2$ the bank cannot serve all remaining depositors. If the bank becomes illiquid, its remaining liabilities are $D-C$. The bank is then placed into receivership, and the value of the loan portfolio at $t=2$ becomes $(1-\rho)\theta L$. For a run not to be an equilibrium, the bank must have enough funds to pay all remaining liabilities at $t=2$. This is the case if $(1-\rho)\theta L \geq D - C$. Hence, there exists a threshold $\theta^{Liquidity}$:
\[\theta^{Liquidity} \equiv \frac{D-C}{(1-\rho)L}> \theta^{Solvency}  =\frac{D-C}{L} \]
such that for $\theta\in[\theta^{Solvency},\theta^{Liquidity})$, self-fulfilling runs constitute an equilibrium. If a bank fails with $\theta$ in this range, the panic run itself is the cause of failure as the bank would have survived absent withdrawals, whereas for $\theta<\theta^{Solvency}$, the bank run is merely the consequence of imminent failure.

\paragraph*{Empirical Predictions}  We next derive empirical predictions from this illustrative framework. The first insight is that $\theta$ must be sufficiently low for a bank failure to occur. If $\theta<\theta^{Solvency}$, a run and failure are unavoidable. If fundamentals are very strong, $\theta>\theta^{Liquidity}$, the bank never fails.

In the intermediate range, $\theta\in[\theta^{Solvency},\theta^{Liquidity})$, there are two pure-strategy equilibria: one where a bank run occurs and one where there is no run. This does not provide a sharp empirical prediction. The literature provides two approaches to pin down the probability of failure.

The first approach assumes agents coordinate on the realization of a publicly observed random ``sunspot'' variable \citep[see, e.g.,][]{Diamond1983,ColeKehoe2000,PeckShell2003,Ennis2009}. The probability of the sunspot could be constant, so that the likelihood of a run is the same for weak and strong banks in this range. Alternatively, the probability could be a function of $\theta$. Perhaps most realistically, it could be declining in $\theta$, so runs are more likely in weaker banks \citep[see, e.g.,][]{Ennis2003,Gertler2015}.

A second approach pins down a unique equilibrium by relaxing the common knowledge assumption in a global game \citep{Morris1998,Morris2003,Rochet2004,Goldstein2005}. In this setting, each agent receives a slightly noisy private signal of $\theta$. This results in a unique threshold equilibrium where all agents withdraw from a bank when the aggregate return on the bank's assets falls short of a cutoff $\theta^*\in (\theta^{Solvency},\theta^{Liquidity})$.\footnote{To yield a run threshold $\theta^* \in (\theta^{Solvency},\theta^{Liquidity})$, the model would require a payoff difference between rolling over and withdrawing in the state where the bank does not fail, such as a positive interest rate paid at $t=2$.}  Importantly, this implies the occurrence of \textit{panic} runs that cause the failure of banks that would have survived absent the run when $\theta \in (\theta^{Solvency},\theta^*)$. The run threshold $\theta^*$ depends on parameters, such as $\rho$, that do not affect fundamental solvency but do affect the degree of strategic complementarity among depositors.

We summarize this discussion with our first prediction:
\begin{hypothesis}
Bank failures are more likely among banks with weak fundamentals (low $\theta$).
\end{hypothesis}
\noindent Note that since $D-C=L-E$, both $\theta^{Solvency}$ and $\theta^{Liquidity}$ are decreasing in the capital ratio $E/L$. Hence, higher capital ratios reduce the likelihood of fundamental and run-triggered failures.

A related empirical prediction concerns the recovery rate for depositors who did not withdraw before failure and thus hold claims in receivership. The pro rata share paid out to depositors in receivership is given by $(1- \rho)\theta L/(D-C)$, which is increasing in $\theta$.
\begin{hypothesis}
    Recovery rates for depositors are increasing in bank fundamentals $\theta$ and are thus lower in fundamentally insolvent banks than in panic-induced failures.
\end{hypothesis}
\noindent If $\rho$ is known, the recovery rate in receivership can allow one to infer information about the realization of $\theta$ and thus whether the bank was fundamentally insolvent.

We consider two extensions of the model to derive predictions on the roles of interbank markets and sleepy depositors. First, to illustrate the role of interbank markets, suppose that at $t=1$ the bank can raise additional cash $Z$  through wholesale or interbank funding. We assume that wholesale funding is always fully secured and available at zero interest. We further assume that the bank can borrow at most $Z\leq \lambda \theta L$, where $\lambda \in [1-\rho,1]$ captures frictions in the interbank market. $\lambda=1$ represents a frictionless interbank market.

With a deep interbank market, the threshold where self-fulfilling runs become possible is lower: $ \theta^{Liquidity,Interbank}\equiv \frac{D-C}{\lambda L} \leq \theta^{Liquidity}$.
Moreover, the range of fundamentals for which runs can cause insolvency is shrinking in the depth of the interbank market, $\lambda$. In the extreme case where $\lambda=1$, self-fulfilling runs never constitute an equilibrium, although fundamental runs can occur ($\theta<\theta^{Solvency}$).

 \begin{hypothesis}
    Interbank liquidity provision reduces the scope for runs to cause bank failures.
\end{hypothesis}

Finally, we consider a variant where some depositors are ``sleepy'' and never withdraw their funds. Let $\overline{D}$ denote the depositors who never withdraw. If sufficiently many depositors are sleepy such that $D-\overline{D}\leq C$, then the bank avoids receivership even if all run-prone deposits withdraw. In this case, even if $\theta<\theta^{Solvency}$, the bank is not closed by a run in $t=1$ and fails only in $t=2$. Bank failure at $t=2$ is predictable in $t=1$. More generally, a larger mass of sleepy depositors weakens strategic complementarities among run-prone depositors and reduces the scope for self-fulfilling runs to occur.

\begin{hypothesis}
    If enough depositors are sleepy, banks can be insolvent but liquid, increasing the predictability of failure.
\end{hypothesis}

\section{Fundamentals and Bank Failures}

\label{sec:evidence}

In theory, failures can be caused by either illiquidity or insolvency. Which type of failure is most empirically relevant? We next discuss the evidence on the causes of bank failures. We organize the evidence around key themes and their implications through the lens of the conceptual framework discussed in \Cref{sec:conceptual}.

We start with Hypothesis 1 from \Cref{sec:conceptual}: bank runs and failures are more likely when fundamentals are weak. The empirical evidence resoundingly backs this prediction. While $\theta$ is typically not directly observable to econometricians, studies across various institutional settings have shown that observable proxies of worse bank fundamentals are associated with a higher chance of bank failure.

\subsection{Aggregate and Narrative Evidence}

The empirical debate on whether bank failures primarily stem from illiquidity or poor fundamentals that trigger insolvency goes back at least to the discussion of the causes of the Great Depression. The Great Depression featured roughly 9,000 bank failures, the largest wave of failures in U.S. history (see \Cref{fig:failures}). In their classic \textit{A Monetary History of the United States}, \citet{FriedmanSchwartz} argued that many bank failures in the Depression resulted from ``self-justifying'' runs that brought down illiquid yet solvent banks (i.e., $\theta > \theta^{Solvency}$). For instance, Milton Friedman famously argued that the failure of the Bank of the United States in December 1930, the largest bank failure in history at the time, was caused by a run that brought down a ``perfectly good bank'' \citep{Friedman1980_FreeToChoose_Ep3}. In Friedman and Schwartz's narrative, the failure of the Bank of the United States and the resulting loss of confidence in the banking system was a key turning point in worsening the Great Depression.

Since \citet{FriedmanSchwartz}, empirical researchers have produced substantial additional evidence on the determinants of bank failures, for the Great Depression and more broadly. Overall, this work has placed more emphasis on poor fundamentals than do \citet{FriedmanSchwartz}. A range of studies using regional data support the intuitive view that bank failures become more likely as fundamentals are weaker \citep{Temin1976,Alston1994,Wicker1996,Wicker2006}. For example,  \citet{Temin1976} estimates regressions explaining state-level bank failures and argues that bank failures in the Great Depression were largely caused by falling income and asset prices from the contraction, rather than by exogenous panics. \citet{Wicker2006} also emphasizes the regional nature of runs and failures in the Great Depression. He argues that runs and failures were generally concentrated in weak banks, rather than following from indiscriminate panics.

\subsection{Bank-level Evidence from Specific Episodes}

Aggregate and regional data have the limitation that they do not reveal whether individual banks that were most likely to fail were weaker on fundamentals. The first systematic studies on historical U.S. bank failures using bank-level micro data are by \citet{White1984}, \cite{Calomiris1997}, and \citet{Calomiris2003a}, which all focus on the Great Depression. \citet{White1984} shows that banks that failed during the 1930 banking crisis were less well capitalized than surviving banks, held more illiquid assets, and relied more on wholesale funding. Failures did not result from a discrete panic but from a continuation of fundamental weaknesses in the banking system, rooted in bad agricultural loans and undiversified portfolios.

Focusing on the June 1932 Chicago banking panic,
\citet{Calomiris1997} find that banks that failed were weaker than surviving banks in terms of market equity values and balance sheet metrics. Failing banks were also more reliant on high-cost borrowing, even before the panic. While there were runs on both solvent and insolvent banks, they conclude that the general runs did not cause the failure of solvent banks in this episode. \citet{Calomiris2003a} document that most banks that failed throughout the Great Depression exhibited observable weaknesses in their balance sheets. Banking panics primarily accelerated the timing of failures among already weak banks, rather than causing the widespread collapse of otherwise solvent institutions.

In the specific case of the Bank of the United States, invoked prominently in \cite{FriedmanSchwartz}, the contention that it was healthy and solvent was later challenged by other scholars \citep{Temin1976,Lucia1985,Obrien1992}. These studies noted that the bank had expanded rapidly in the 1920s, had unusually large real estate loan exposure in a declining real estate market, and was engaged in inside lending and fraudulent deception to hide the bank's weaknesses. Bank examiners were aware of these issues well in advance of its failure in December 1930.

Research on more recent episodes also finds that failures are strongly related to weak bank fundamentals. Studies on the 1980s Savings and Loan Crisis and the 2008 Global Financial Crisis (GFC) systematically find that banks that are highly levered, have low earnings and liquidity, and hold risky asset portfolios are more likely to fail \citep[][]{Cole1995,Cole1998,Wheelock2000,BergerBouwman2013}. The finding that poor fundamentals are necessary for failure in modern times is less surprising, given that deposit insurance reduces the scope for runs to cause failure. Nevertheless, the share of uninsured deposits has risen considerably in recent years \citep{HansonBPEA2024}. The runs on Silicon Valley Bank and other banks in March 2023 showed that runs are not a relic of history. Consistent with the historical evidence, the banks that failed in March 2023 had also suffered large asset losses, in this case on long-term securities holdings, pushing them toward insolvency \citep{Jiang2023,Metrick2023}.

\subsection{Broader Bank-Level Evidence on Fundamentals and Bank Failures}

Evidence from specific episodes like the Great Depression suggests that weak fundamentals are necessary for explaining bank failures. In \citet*{Correia2025FailingBanks}, we generalize this insight. We bring together new systematic evidence from 160 years of micro-level U.S. data covering 1863--2024 based on a major data collection effort. Studying such a long sample has several benefits. First, it allows for the study of a much larger number of events, covering 5,120 bank failures.\footnote{The data in \citet{Correia2025FailingBanks} consist of a historical sample that covers all national banks from 1863 to 1941 and a modern sample that covers all commercial banks from 1959 to 2024. The historical U.S. banking system featured banks chartered under both state and federal law. The focus on national banks chartered under federal law is dictated by the availability of consistent data.} Second, this long sample spans a range of institutional and regulatory regimes, covering periods both with and without deposit insurance and a public lender of last resort. Finally, it covers both crisis times and normal times, which provides insights into whether bank failures during crises are different and whether crises are predictable.

In \citet*{Correia2025FailingBanks}, we establish that observable weak fundamentals systematically precede bank failure. Throughout the sample, failing banks are characterized by declining net income and capitalization, rising asset losses, an increased reliance on noncore funding, and an asset boom and bust.  This holds across various institutional settings, including and excluding deposit insurance and a public lender of last resort. Failures during major financial crises are similarly linked to weak fundamentals as failures during quiet periods.

\begin{figure}[h!]
\subfloat[Surplus profit relative to total equity]{\includegraphics[width=0.49\textwidth]{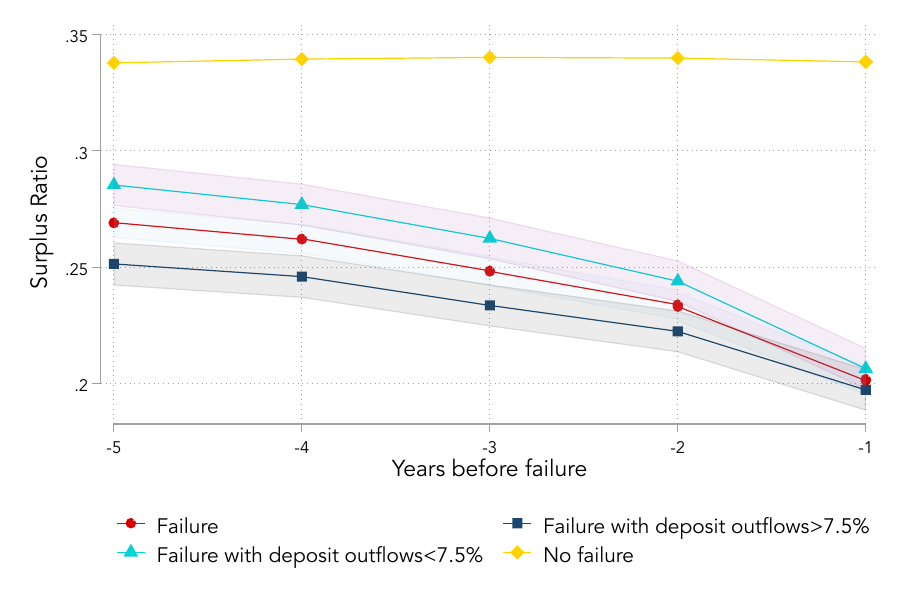}}
\hfill
\subfloat[Nonperforming loans to total loans ]{\includegraphics[width=0.49\textwidth]{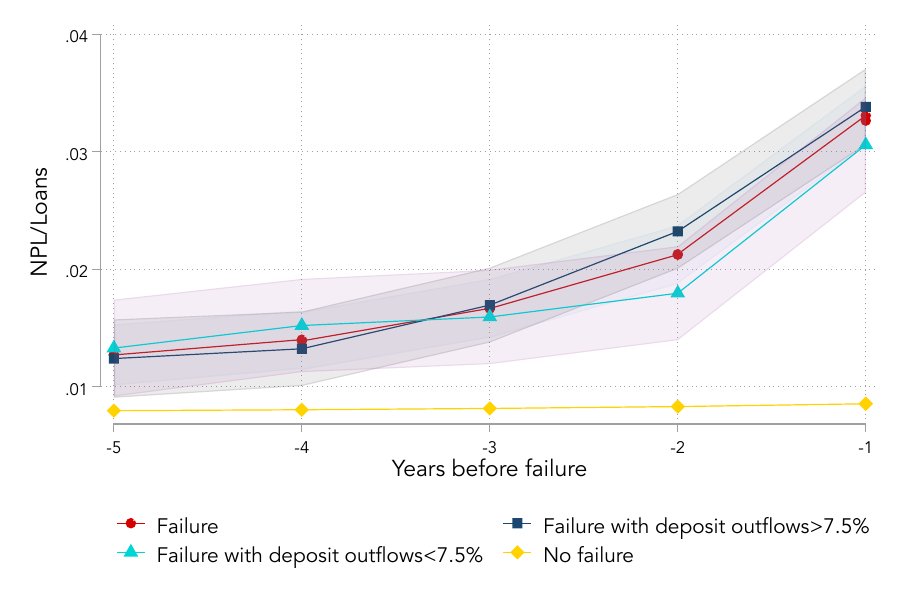}}

\subfloat[Noncore funding ratio]{\includegraphics[width=0.49\textwidth]{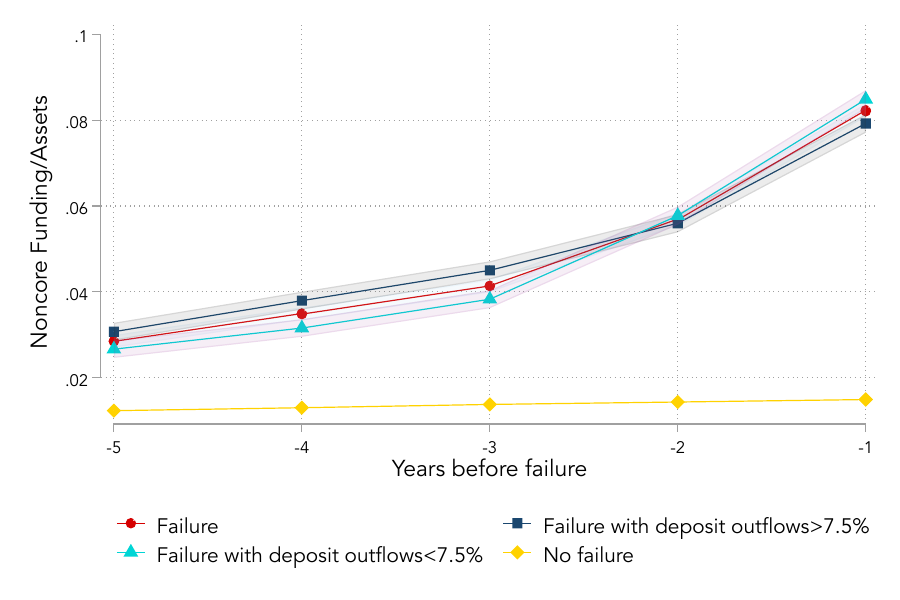}}
\hfill
\subfloat[Deposits to assets]{\includegraphics[width=0.49\textwidth]{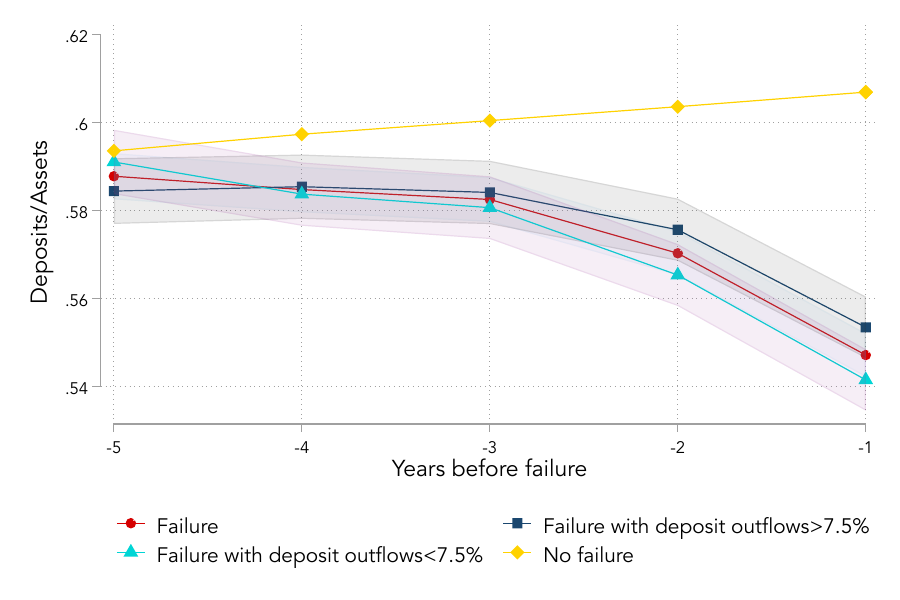}}

\vspace{0.5cm}
\caption{Failing Banks Exhibit Deteriorating Fundamentals Irrespective of a Run}
\label{fig:failure_fundamentals}
\begin{tabnote}
Notes: This figure shows coefficients obtained from estimating a regression of bank financial ratios on three mutually exclusive indicator variables, while controlling for year fixed effects. The three regressors are (i) bank fails with a run, defined as deposit outflows of more than 7.5\%, (ii) bank fails without a run, and (iii) the bank does not fail in year $t$. We also report the coefficient for all failures from a separate specification that combines (i) and (ii). We exclude the constant term, so the estimated coefficient is the average level of $y_{bt+h}$ before failure. Time $t=-1$ is the last call report before failure. We focus on the sample of all national banks from 1863 to 1934. We also include confidence intervals for the three sets of failure coefficients in order to assess whether they were statistically different from the no-failure coefficients.
\end{tabnote}
\end{figure}

\Cref{fig:failure_fundamentals} illustrates the deterioration in fundamentals in the run-up to failure. The analysis is based on all national banks from 1863 to 1934, a sample that spans the start of the National Banking Era through the Great Depression and the introduction of deposit insurance. It plots the average level of various bank fundamentals for 2,887 failing banks and for surviving banks. Even five years before failure, failing banks on average display a significantly worse position in terms of capitalization and recent retained earnings, nonperforming loans, and reliance on expensive noncore funding.\footnote{While bank fundamentals are most commonly proxied with direct measures of insolvency risk such as net income, capitalization, or asset losses, several studies highlight that bank fundamentals can also be proxied with measures of noncore funding such as time deposits or other forms of wholesale funding \citep{White1984,Calomiris2003a,Correia2025FailingBanks}. Banks subject to losses often were forced to rely on noncore funding. This, in turn, further exacerbated losses, as this form of funding was typically more expensive and risk-sensitive. Reliance on noncore funding may, in principle, also make banks more exposed to funding shocks. However, a higher reliance on noncore funding in a failing bank does not imply that a run was more likely to be the cause of failure.} Moreover, failing banks see a gradual deterioration in each of these metrics in the five years before failure. Even though banks did not face strict provisioning requirements under historical accounting standards, financial statements still reveal significant weaknesses in failing banks in the years before failure. Failing banks also see a gradual loss in deposits, another indication of a deteriorating business.

In \cite*{Correia2025FailingBanks}, we further show that bank failures are substantially \textit{predictable} out-of-sample, both in the historical sample before deposit insurance and in the modern banking system. Aggregate waves of failures during banking crises are also predictable using micro-data on poor bank fundamentals. Thus, banking crises with many failures should not be seen merely as unexpected jumps to bad equilibria; they largely reflect poor fundamentals. Moreover, the substantial predictability of bank failures indicates that depositors were often slow to respond to bank weakness.  This suggests a potentially important role for sleepy depositors who allow banks to be insolvent but remain liquid for a time, making failures more predictable, as outlined in Hypothesis 4 of \Cref{sec:conceptual} \citep[see also,][]{HANSON2015449,Jiang2023}.

\subsection{Fundamentals in Failures With and Without Runs}

A necessary, but not sufficient, condition for a run to cause failure of a solvent bank is for deposits to actually flow out of the bank, forcing it to undertake value-destroying actions. In \citet{Correia2025FailingBanks}, we examine deposit outflows immediately before failure for all national bank failures from 1880 to 1934 and for all commercial bank failures from 1992 to 2024. Before the introduction of deposit insurance, the average failing bank lost 14\% of its deposits before failure, and one-quarter of failures involved deposit outflows greater than 20\%. Runs in failing banks were thus common before deposit insurance, as illustrated in \Cref{fig:failures}.  In contrast, runs are much less common in modern-day bank failures. After the founding of the FDIC, failing banks only lose around 2\% of their deposits before failure.

\Cref{fig:failure_fundamentals} separates failures into those with and without runs, defined as a deposit outflow immediately before failure exceeding 7.5\%. A striking finding in \Cref{fig:failure_fundamentals} is that the deterioration in fundamentals is similar for failures with and without runs.  Failures with runs are not less tightly linked to poor fundamentals. Runs do not appear to trigger the failure of relatively healthier banks. Rather, this suggests that runs trigger the failure of banks that likely would have failed even absent a run. As a result, failures with large deposit outflows are roughly as predictable as other failures \citep{Correia2025FailingBanks}.

The evidence that failures with runs are strongly related to weak fundamentals argues against the importance of random, non-fundamental runs of the sort invoked by \cite{FriedmanSchwartz} in explaining many bank failures. However, it does not preclude the importance of fundamental-based panic runs in explaining failures. Moreover, the finding that non-fundamental runs are a rare cause of failure does not mean that runs on healthy banks did not occur; these runs just rarely resulted in bank failure.

\subsection{Weak Fundamentals and Bank Asset Booms}

What causes weak bank fundamentals? Unexpected asset shocks, such as crop failures, real estate price declines, failure of a major borrower, or a local economic downturn, are typical drivers. In addition, the literature has identified another important precursor of poor fundamentals: rapid asset growth, usually through rapid lending growth or acquisitions. Rapid loan growth is often associated with rising bank leverage and future loan losses. Most of the major bank failures throughout U.S. history occurred after rapid, and arguably reckless, loan expansion. Examples include the already-mentioned failure of the Bank of the United States in 1930, Franklin National Bank in 1974, Continental Illinois in 1984, and Washington Mutual in 2008, all of which were the largest bank failures in U.S. history at the time.

Consistent with this, a large body of evidence using aggregate data finds that rapid expansion in credit is a strong predictor of banking crises \citep{Schularick2012,Baron2017,Greenwood2022,MullerVerner2023}. Bank-level evidence also confirms the importance of rapid bank asset growth in presaging subsequent losses. \cite*{Fahlenbrach2018} use data from 1973 to 2014 and find that banks with rapid asset growth have lower future stock returns. \cite*{Correia2025FailingBanks} show that rapid loan growth is a strong predictor of bank failure in the next three-to-five years, especially in the post-WWII sample, when banks were more unconstrained in their ability to grow. This finding suggests that neglect of downside risks or governance problems can lead banks to take on excessive risks and grow too quickly when times are good, systematically leading to higher losses in bad times \citep{Bordalo2018,Greenwood2024reflexivity}.

\section{Recovery Rates in Failure}

The evidence reviewed so far overwhelmingly supports a link between bank failures and weak fundamentals. At the same time, pre-FDIC failures were commonly associated with runs. The evidence on weak fundamentals is not conclusive about whether failures primarily reflected runs on weak but solvent banks ($\theta^{Solvency}<\theta<\theta^{Liquidity}$) or on fundamentally insolvent banks ($\theta<\theta^{Solvency}$).

Ascertaining whether a bank is fundamentally insolvent or weak but solvent is empirically challenging. Bank assets are opaque and hard to value. This is especially true in a crisis and in historical accounting data, as banks were not required to provision for losses. The framework in~\Cref{sec:conceptual} suggests that one approach to make progress on understanding this challenging question is  to study recovery rates in failure. Specifically, Hypothesis~2 predicts that recovery rates for creditors are lower in fundamentally insolvent banks.

In \citet*{Correia2025FailingBanks}, we exploit this insight using detailed information on receivership proceedings for all national banks failures from 1865 to 1934. A key challenge for interpreting recovery rates is that failure itself can reduce the value of bank assets. Asset payoffs may be tied to bankers' human capital and can thus be lower in receivership than in the hands of the banker. In our model, asset values fall by $\rho$ in receivership, but $\rho$ is unobservable. Therefore, using recovery rates to infer fundamental insolvency requires a prior on the value destroyed by failure.\footnote{As discussed in more detail by \citet{Correia2025FailingBanks}, beyond receivership asset losses, failure could also destroy bank franchise value \citep[see, e.g.,][]{DSSW2023} that is not reflected by the assets held on the bank's balance sheet. Thus, using recovery rates to understand the causes of failure requires a prior on both receivership asset losses and the franchise-value loss.}

In the context of national bank receiverships, receivers understood that fire sales of failed-bank assets were detrimental to recovery~\citep{ContiBrownVanatta2025}. They were required to liquidate assets in an orderly fashion and could not sell them without a court order and the approval of the Comptroller of the Currency. More broadly, receivers were considered experts at collecting failed banks' assets~\citep{upham1934closed}.  Assets were typically held and collected over several years, and there is ample evidence that the Office of the Comptroller of the Currency (OCC) avoided fire sales into illiquid markets. Thus, the loss due to receivership is not necessarily large, especially considering that the data suggest receivers often replaced bankers managing gradually failing businesses.

The data on national bank receiverships reveal that recovery rates on assets fell substantially short of creditor claims.  \Cref{tab:recovery_rates} reports statistics on the ratio of funds collected from assets in receivership to outstanding debt claims at failure. This ratio represents the average creditor recovery rate, an indication of the extent to which the bank was underwater on its debts. Asset recovery, on average, covered only 75\% of debt claims.\footnote{Note that the creditor recovery rate differs from the asset recovery rate. The latter is defined as funds collected from assets relative to the nominal value of assets at failure. \citet{Correia2025FailingBanks} show that the average asset recovery rate in failed banks for this sample was 51\%. } Moreover, creditors experienced positive losses in 81\% of failures. \Cref{tab:recovery_rates} shows that recovery rates were even lower in failed banks with larger pre-failure deposit outflows. Asset recovery relative to debt in these failures averaged 72\%, consistent with runs occurring in the weakest banks. These low recoveries in pre-FDIC failures reinforce the view that poor asset quality was central to bank failures. We note that these calculations do not account for the fact that depositors were generally paid over several years. Accounting for the opportunity cost of funds would imply even lower creditor recovery rates.

\begin{table}
\maybescriptsize
\setlength{\tabcolsep}{2.5pt}
\caption{Creditor Recovery Rates and Asset Quality in National Bank Failures, 1865--1934}
\label{tab:recovery_rates}
\begin{center}
\begin{tabular}{lcccccc|cccc}
\toprule
Sample  & \multicolumn{6}{c}{Creditor recovery rate}& \multicolumn{3}{c}{Asset assessments}  & $N$ \\
\cmidrule(lr){2-7}\cmidrule(lr){8-10}

 & Avg. & $<$50\%  & [50,70\%)  & [70,85\%)& [85,100\%) &  $\geq$100\% &  Good & Doubtful & Worthless  & \\ \midrule
All&0.75&0.16&0.20&0.22&0.21&0.19&0.36&0.47&0.18&2935\\
   \\

Outflow: & & & & & & & & & \\
Outflow < 7.5\%&0.83&0.09&0.19&0.21&0.29&0.23&0.40&0.45&0.15&1153\\
Outflow > 7.5\%&0.72&0.20&0.23&0.26&0.19&0.12&0.33&0.49&0.20&1599\\
  \\
\bottomrule
\end{tabular}
\end{center}
\begin{tabnote}

Notes:  This table reports (i) estimated creditor recovery rates and (ii) the shares of assets assessed as good, doubtful, and worthless at suspension for failed banks.  The creditor recovery rate is defined as the ratio of funds collected from assets in receivership relative to estimated total outstanding debt claims at suspension. This represents the average recovery rate for all debt claim holders. Asset assessments are based on OCC estimates of the composition of good, doubtful, and worthless assets at suspension. Good, doubtful, and worthless assets at suspension are normalized by total assets at suspension. The sample covers failed national banks from 1865 to 1934. Data are collected from the OCC Annual Report to Congress, tables on ``National banks in charge of receivers'' (various years). In the second and third rows, we split the sample into failed banks with deposit outflows of more and less 7.5\% between the last call report and the time of bank suspension. Data on deposit outflows are only available for the 1880--1934 subsample.
\end{tabnote}
\end{table}

Low recovery rates imply that most, but not all, banks that experienced a run and failed in the pre-FDIC era were fundamentally insolvent ($\theta<\theta^{Solvency}$), unless one assumes large value destruction from failure itself.  For a run to plausibly cause the failure of a solvent bank, two conditions must be met. First, the bank must experience a substantial deposit outflow before failure. Second, recovery rates must imply that the bank would have been solvent outside of failure, conditional on an assumption about value destruction of failure. In \citet*{Correia2025FailingBanks} we calculate that, under the extreme assumption of no failure-induced value destruction, 8\% of pre-FDIC failures involved a run on a fundamentally solvent bank. Under an arguably equally extreme assumption that failure destroys 20\% of bank value, the share rises to 22\%. Thus, even before deposit insurance, runs on weak but fundamentally solvent banks likely accounted for a modest, though not negligible, share of failures.

\section{Expert Assessments of the Causes of Bank Failures}

Subjective assessments of the causes of bank failures by contemporary bank examiners and other experts provide additional insights into the roles of solvency and liquidity. As early as the mid-19th century, bank examiners were concerned with assessing asset quality and understanding the causes of individual bank failures \citep{ContiBrownVanatta2025}. These expert assessments provide a valuable complement to financial data.

For all failed banks from 1865 through 1941, the OCC provided examiner assessments of bank asset quality at the time of suspension. Examiners grouped assets into three categories: good, doubtful, and worthless. These assessments are highly predictive of recovery rates \citep*{Correia2025FailingBanks}. A regression of the asset recovery rate on the shares of good, doubtful, and worthless assets has an $R^2$ of 94\%. Each dollar of good, doubtful, and worthless assets yielded 89, 54, and 8 cents in recovery, respectively.

As shown in \Cref{tab:recovery_rates}, for the average failed bank, OCC examiners assessed only 36\% of failed bank assets as good, 47\% as doubtful, and 18\% as worthless. On-the-ground examiners inspecting bank assets believed that the typical failed bank held highly troubled assets, consistent with the low subsequent recovery on these assets. This holds both for banks subject to runs and for those that failed without a run. In fact, banks subject to runs were assessed as having slightly lower-quality assets than other failed banks.

Moreover, for failures from 1865 through 1928, the OCC systematically provided bank-specific assessments of the cause of failure. Focusing on six panic years during the National Banking Era, \cite{Calomiris1991} highlight that in those panic years only a single bank failure was classified by the OCC as having been caused by a run. \citet{Correia2025FailingBanks} extend this analysis and provide a comprehensive study of the OCC's cause of failures for all failures through 1928 (with partial coverage for 1929--1939). The most common causes of failure are poor local economic conditions, losses, and fraud. Bank runs and illiquidity are cited as the cause of bank failure in fewer than 20 out of over 2,000 failures for which a cause is provided.

A limitation of the OCC cause-of-failure data is its incomplete coverage for failures during the Great Depression. \citet{Richardson2007} studies the classifications of bank suspensions from the Federal Reserve Board between 1929 and 1933, providing a more complete picture of examiner assessments of bank failure during the Great Depression. These classifications indicate that asset losses were the primary source of bank distress. However, they also suggest that illiquidity may have played a larger role in bank failures during the Great Depression than during most other periods in U.S. history. The notion that the Depression was ``different'' is also supported by econometric evidence in \citet*{Correia2025FailingBanks}. They document an increase in excess failures at the height of the Depression, beyond what \textit{ex ante} fundamentals can account for, potentially consistent with important roles for contagion and amplification through asset price declines. Deposit outflows at failing banks were larger during the Depression, suggesting that many more failures involved runs \citep{Correia2025FailingBanks}. At the same time, during the Depression, OCC examiner assessments of asset quality and recovery rates were very pessimistic, suggesting that poor asset quality remained the root source of distress. Consistent with this, \cite{Calomiris1997}  conduct a detailed analysis of OCC examiner reports for banks that failed during the 1932 Chicago panic and find that failed banks were known to have deeply troubled assets before failure.

Broader studies on the causes of bank failures by regulators describe most failures as stemming from asset losses (in line with $\theta<\theta^{Solvency}$). A report by the \citet{Fed1936} analyzing bank suspensions from 1892 through 1935 argues: ``In our long, failure-studded history of banking most of the institutions which suspended business were subsequently proved to be insolvent.'' The report goes on to argue that the immediate trigger of failure is typically a lack of cash, in many cases from deposit outflows, but that this is not the root cause:
\begin{quote}
\textit{While the loss of cash reserves is the immediate cause of the majority of suspensions it is not the fundamental or underlying cause. The loss of cash is something that can happen to almost any bank, and by the tenets of sound banking this contingency should be provided for in the loan, investment, and reserve policies. The inability to replenish cash reserves is a condition which arises from holding assets of an inferior quality---assets which cannot be sold without loss or used as collateral for borrowing.}
\end{quote}

In the contemporary era, \citet{OCC1988} study 171 bank failures from 1979 to 1987. They argue that most failures in their sample were due to poor management practices that resulted in weak loan policies, excessive risk-taking, and insider abuse. These weaknesses were often exacerbated, but seldom solely caused, by economic decline.

\section{Why Are Runs on Solvent Banks a Rare Cause of Failure?}

A large body of empirical evidence indicates that fundamental insolvency, rather than illiquidity or depositor runs, is the dominant driver of bank failures. Runs can be an important trigger for the failure of insolvent institutions, but runs less commonly cause the failure of fundamentally sound banks. This raises the question: Why do runs rarely cause solvent banks to fail?

\subsection{Runs and Bank Fundamentals}

One possible explanation is that only insolvent banks are subject to runs. This hypothesis, however, is not supported by the data. While the evidence suggests that runs are more likely in weak banks and after bad economic news, healthy banks can also be subject to runs.

Using aggregate time-series data, \cite{Gorton1988} and \cite{Calomiris1991} argue that U.S. banking panics during the National Banking Era were predictable reactions to fundamental shocks that altered depositor perceptions of bank solvency. Using monthly data from banking crises in 46 countries since 1870, \cite*{Baron2021} find that banking panics were preceded by substantial declines in bank stocks, suggesting that panics occur following the realization of weak fundamentals that lower bank equity values. \cite{JamilovKoenigMuellerSaidi2025} use a narrative approach in a cross-country panel to document that most, but not all, systemic bank runs are triggered by adverse news about banking-sector or macroeconomic fundamentals.

Micro-level evidence also supports the notion that runs are more likely after bad news about bank fundamentals. \citet{Correia2025bankruns} use textual analysis of newspaper articles to identify over 4,000 runs on U.S. banks from 1863 to 1934. They find that, while runs were more likely in banks with weaker fundamentals, strong banks were also subject to runs, often in response to negative news about the banking system or economy. However, strong banks rarely failed following a run.

The Global Financial Crisis (GFC) also saw runs, although mostly outside of the regulated banking sector. Rollover freezes occurred in the asset-backed commercial paper (ABCP) market \citep[e.g.,][]{Covitz2013} and the bilateral repo market  \citep[e.g.,][]{Gorton2012,Krishnamurthy2014}. Moreover, the GFC also featured counterparty runs on the prominent investment banks Bear Stearns and Lehman Brothers \citep[see, e.g.,][]{Copeland2014} and a run on prime money market funds following the failure of Lehman \citep{Schmidt2016}. However, in retrospect,  the panics of the GFC seem anything but random. Key events in the GFC were directly related to the collapse of the housing market and resulting asset losses \citep[see, e.g.,][]{Gennaioli2018}.

Research on other episodes confirms that runs are more likely at weak banks, but not always confined to them. \cite{Calomiris1997} finds that both solvent and weak banks experienced runs during the Chicago Panic of 1932, but only insolvent banks failed. \citet{Schumacher2000}  studies the runs on Argentina's banks following Mexico's devaluation in late 1994. Bank runs were driven by informed withdrawals on weaker banks whose solvency was more likely to be impaired by the shock. \cite{CiprianiEtAl2024} document runs on 22 banks during the 2023 U.S. banking turmoil. They also find that runs are more likely at banks with weak fundamentals, both in terms of solvency but also in terms of liquidity and other characteristics such as having more granular deposits and a publicly observable stock price, consistent with a global game model.

The evidence that runs can afflict both weak and strong banks during panics is consistent with asymmetric information theories of panics \citep{Jacklin1988,Chari1988,Calomiris1991}. When depositors observe a public signal that some banks might be insolvent but cannot fully distinguish weak from strong banks, they run on both healthy and weak banks \citep{Dang2020information}. Strong banks can also be subject to runs due to false rumors. Evidence from \cite{Correia2025bankruns} points to a role for frictions in information processing, as bank fundamentals provide substantially greater ability to distinguish weak from healthy banks than do observed runs.

Strategic complementarities play a central role in generating runs on fundamentally solvent banks. For weak but solvent banks, depositors' incentive to run is increasing in their belief that others will run. While estimating role for strategic complementarities is empirically challenging, existing studies find an increased propensity to run in contexts where strategic complementarities are likely stronger. The framework in \Cref{sec:conceptual} implies that strategic complementarities are stronger when assets are less liquid or subject to a larger loss in receivership ($\rho$ is larger). Along these lines,  \cite{ChenGoldsteinJiang2010} present evidence for the role of strategic complementarities in US equity mutual funds, showing that outflows are more sensitive to past performance in funds holding illiquid assets, where strategic complementarities are expected to be stronger. For U.S. banks,  \cite{chen2024liquidity} document that uninsured deposit outflows are more sensitive to past performance in banks with greater liquidity mismatch. \citet{Artavanis2022} exploit a policy uncertainty shock during the aftermath of the 2014--2015 Greek sovereign debt crisis. Using granular depositor-level data, they estimate that two-thirds of deposit withdrawals were driven by deteriorating fundamentals, while the remainder was due to strategic complementarities.

\subsection{Mechanisms for Resolving Runs and Depositor Heterogeneity}

The reason runs do not trigger failures of healthy banks is not that such runs do not occur. Rather, runs are often resolved by mechanisms other than fire sales, preventing inefficient failures of solvent banks, even absent government intervention.

\cite{Correia2025bankruns} analyze newspaper descriptions of how banks in the U.S. responded to runs prior to the introduction of deposit insurance. Banks that survive runs are often reported to accommodate withdrawals to calm depositors. At the same time, banks often attempt to restore confidence by signaling strength, such as by conspicuously delivering ``truckloads of cash.'' Cash could come from a new equity or debt injection from bank owners or other investors.

In the historical U.S. banking system, more severe runs would lead banks to suspended convertibility, either partially or fully  \citep{Sprague1910,Gorton1985suspension}. Suspension was often undertaken jointly through local clearinghouse associations. Suspension allowed the clearinghouse and bank examiners to audit distressed banks to assess the state of bank investments and verify solvency. Banks that were deemed solvent were reopened. While suspensions were not legally authorized \citep{Gorton1985suspension}, bank examiners allowed and even encouraged this practice when they believed funding problems were temporary \citep{ContiBrownVanatta2025}. Suspensions helped prevent unwarranted failures of solvent banks, but they could nevertheless lead to temporary disruptions for the local economy \citep{Gorton2012book}.

Interbank arrangements can also allow a solvent bank to weather a run. The model in \Cref{sec:conceptual} shows that a well established interbank market reduces the scope for self-fulfilling runs to cause the failure of solvent banks (Hypothesis 3). In this case, solvent banks can borrow from other banks in response to deposit withdrawals. Thus, when interbank participants can discern whether distress stems from liquidity rather than solvency concerns, interbank lending can insure against temporary funding pressures from runs.

In the era before the Federal Reserve, clearinghouses would act as quasi-central banks by issuing loan certificates, a joint liability of all members, to provide liquidity \citep{Timberlake1984}. In 1893 and 1907, clearinghouses issued small-denomination certificates directly to the public. Suspension and cooperation through clearinghouses could avoid destructive asset fire sales. The NYCH performed these functions effectively in the Panic of 1873, but not in later panics because member banks struggled to internalize the collective benefits of such interventions \citep{Wicker2006}.

Evidence suggests that interbank markets and other informed depositors can often discriminate between weak and strong banks, mitigating the adverse consequences of runs on healthy banks. In a narrative description of runs, \cite{Nicholas1907runs}\footnote{This source is also discussed in \cite{CalomirisKahn1991}.} writes: ``one kind of bank run [is] the descent of a crowd of angry and frightened depositors to withdraw deposits. There is another kind of run, less sensational in appearance, but usually much more serious in its results. If a bank is actually in bad shape there is far more likelihood of its actual condition being discovered by other banking institutions than by the individual depositors of the bank.'' \citet*{afonso2011stressed} show that during the 2008 Global Financial Crisis, the U.S. federal funds market remained active but segmented, with lending continuing mainly between counterparties with established relationships, while riskier borrowers faced sharply higher rates. \citet*{Perignon2018} find that wholesale funding ``dry-ups'' in the European CD market from 2008 to 2014 were bank-specific and driven by informed lenders withdrawing funds from weaker banks, rather than market-wide freezes or adverse-selection exits by strong banks. \citet{Blickle2022} study a system-wide bank run during the German Crisis of 1931, a key event of the Great Depression. While most banks lost deposits during this major run, interbank deposits declined almost exclusively at failing banks. This pattern suggests that banks are better informed about which peer banks will fail.

More broadly, the empirical literature on withdrawals finds substantial heterogeneity in depositors' ability to distinguish between distressed and non-distressed banks. For example, \citet{Currie1938SuspendedBanks} shows that deposit outflows during the Depression were mostly from large depositors, suggesting that retail depositors---who tend to be less informed---were less likely to withdraw from failing banks.  \citet{Saunders1996} find evidence that informed depositors could distinguish between failing and surviving banks during the Depression. \citet{OgradaWhite} study runs on the Emigrant Industrial Savings Bank during the Panics of 1854 and 1857. They find that the bank could withstand an unwarranted run by uninformed depositors in the Panic of 1854, but was forced to suspend convertibility in 1857, when the run was initiated by more informed depositors. \citet{Iyer2012} use granular depositor-level data to study a rumor-based run on an Indian bank and find that the bank survived not because of deposit insurance alone but because of the trust built through long-standing depositor relationships and social networks. Building on this evidence, \citet*{Iyer2016} find that a high-solvency-risk shock led to runs by informed depositors. By contrast, in a rumor-based run, informed depositors stood firm while uninformed depositors drove withdrawals.

Taken together, the evidence across various settings suggests that strong banks can usually survive an unwarranted run, even in the absence of government support, through interbank linkages, relationships with their depositors, signaling strength, or, at worst, suspension.  However, while interbank markets help banks insure against liquidity risk, they can also be a source of contagion \citep[see, e.g.,][]{Allen2000}. For instance,  \citet{Iyer2011} demonstrate that the failure of an Indian bank triggered additional withdrawals at banks exposed to the failing bank. \citet{Mitchener2019} suggest that interbank connections amplified the contraction in lending during the Great Depression.

\section{Government Interventions}

We next consider the two most prominent policies used to prevent or contain bank failures driven by runs: deposit insurance and the lender of last resort.

\subsection{Deposit Insurance and the Role of Runs versus Supervisory Closures}

Deposit insurance crucially shapes the dynamics of bank failures and bank runs. \Cref{fig:failures} shows that, although the FDIC era has witnessed major waves of bank failure, failures involving runs have become much less common.

Deposit insurance has changed how banks fail.  Even if runs on solvent banks were not a common cause of failure before deposit insurance, runs may have been an important trigger for failures of insolvent banks, as in \citet{Diamond2001}. Deposit insurance reduces the scope for these runs that impose discipline on insolvent banks. As a result, as \citet*{Correia2025supervising} show, bank failures in the modern banking system are usually supervisory decisions. Supervisory discipline is valuable in the absence of market discipline, as deposit insurance can allow failing banks to operate longer than optimal. For instance, \citet{PuriJF} show that insured deposits tend to flow into failing banks, offsetting withdrawals from uninsured depositors. Having insolvent banks fail through supervisory closure rather than through runs may reduce disruptions and the costs of failure. However, it also expands the scope for supervisory forbearance and its associated costs~\citep{Kane1989Book,Caballero2008}.

Deposit insurance can also alter banks' risk-taking incentives. \citet{calomiris2019stealing} exploit U.S. state-level experiments in partial deposit insurance before the FDIC. They find that insured banks expanded their deposits and lending during the agricultural boom in WWI, while decreasing cash and capital buffers. Ultimately, the additional risk-taking resulted in steep losses and systemic failures when agricultural prices fell after WWI.

Additional evidence on deposit insurance comes from \citet{Iyer2019} and \citet{CucicIyerKokasPeydroPica2024}, who use granular account-level data to study a temporary shift to unlimited deposit insurance in Denmark during the GFC.
\citet{CucicIyerKokasPeydroPica2024} show that the initial expansion to unlimited deposit insurance induced deposits to flow disproportionately to weaker banks with riskier loan portfolios, allowing them to continue supplying credit to lower-productivity, higher-risk firms. \citet{Iyer2019} find that the shift back to limited deposit insurance disproportionately benefited large, systemically important banks, while smaller banks suffered funding shocks and curtailed lending.

Overall, the evidence suggests that deposit insurance shapes bank funding conditions, although the effect depends on the credibility of the insurance scheme \citep{MartinezPeria2001}. Changes in funding conditions, in turn, may affect banks' investment decisions. The benefits of eliminating possibly destructive bank runs---including those in fundamentally insolvent banks---must be weighed against their costs.

\subsection{Lender of Last Resort}

Our model above suggests that liquidity should be provided to solvent banks subject to runs ($\theta> \theta^{Solvency}$) whenever there is no functioning interbank market. This reflects the logic underlying the famous \citet{Bagehot1873} doctrine. To contain a panic, a central bank should lend freely to solvent institutions against collateral at high rates. What does the existing empirical evidence suggest about the effectiveness of liquidity provision to distressed banks?

\citet{Richardson2009} exploit a natural experiment during the Great Depression based on the differential discount window lending policy of the Atlanta and St. Louis Federal Reserve banks. Banks in the Atlanta Fed's jurisdiction, which championed generous liquidity support, experienced significantly fewer suspensions, better lending conditions, and quicker economic recovery, particularly during the 1930 panic. The evidence supports the notion that central bank liquidity provision effectively mitigated banking panics.

International evidence indicates, however, that preventing panics is insufficient to avoid the adverse consequences of crises. \cite*{Baron2021} use a cross-country panel to study banking crises with and without panics. Even when policies forestall runs, ``quiet crises''---banking sector distress without panics---result in severe contractions of credit and output.  Because banking crises are typically rooted in solvency problems, they cannot be mitigated by liquidity policy alone.  Consistent with this, \citet{Baron2024} show, using data on policy interventions collected by \cite{MetrickSchmelzing2021}, that, while aggressive liquidity interventions can be helpful, they typically produce only modest, short-lived increases in the market value of bank stocks. Liquidity interventions cannot reverse the long-run undercapitalization that follows substantial equity losses.

\section{Avenues for Future Research}

The existing evidence argues for the view that most bank failures are driven by fundamental losses and insolvency. Runs are often the trigger or accelerant of failure for fundamentally insolvent banks. However, runs less commonly cause the failure of otherwise solvent banks, as such runs are often resolved through less costly mechanisms than liquidation. Nonetheless, several important questions remain unanswered. We propose three themes for future research.

First, while there is a large theoretical literature on bank runs, most of this literature does not capture the dynamics of runs and failures. In the data, failing banks exhibit gradually rising losses and deteriorating funding conditions, generating feedback between losses and funding costs. We conjecture that incorporating behavioral frictions is important for accurately modeling these dynamics. Failing banks can often remain liquid even when seemingly insolvent, increasing the predictability of failures and suggesting that some depositors are ``sleepy.'' At the same time, even healthy banks can face runs, but these runs rarely result in failure, suggesting that ``sleepy'' depositors occasionally wake up and run on banks without causing failure. Embedding such frictions into dynamic models of bank runs and bank failures will allow for a better understanding of run dynamics. A few papers have made progress in this direction \citep{LorenzoniWerning2019,LlambiasOrdonez2024}.

Second, too little is known about how banks withstand runs absent government support. Solvent banks can survive runs without deposit insurance or central bank support. While interbank markets and private liquidity arrangements often play a stabilizing role, more work is needed to understand their limits. Under what conditions do market failures prevent liquidity from reaching solvent but distressed institutions? A clearer understanding of these mechanisms is key to identifying the conditions under which various crisis interventions, such as emergency liquidity provision or forced recapitalization, are optimal.

Third, asset losses and realized credit risk are common sources of failure. What conditions encourage the risk-taking that leads to such losses? When does a bank's governance structure allow managers to take risks that ultimately threaten the bank’s survival? And why do some banks fail in predictable ways while others do not? Addressing these questions would deepen our understanding of both the microeconomics of bank failures and the macroeconomics of credit cycles.

\clearpage

{\singlespacing
\footnotesize
\bibliographystyle{chicagofixed}
\bibliography{literature}
}

\end{document}